# EMERGENCE OF CONFORMIST THINKING UNDER STRONG PARADIGMATIC PRESSURE: THE CASE OF (GALILEAN) RELATIVITY

**Estelle Blanquet**

*INSPÉ, Académie de Bordeaux, Université de Bordeaux,
and Laboratoire Cultures — Éducation — Sociétés (LACES, EA7437),
3 ter place de la Victoire, Université de Bordeaux, 33076 Bordeaux, France
estelle.blanquet@u-bordeaux.fr*

**Éric Picholle**

*Institut de Physique de Nice (INPHYNI), UMR 7010 CNRS - Université Côte d'Azur,
Parc Valrose, 06108 Nice cedex, France
eric.picholle@inphyni.cnrs.fr*

(December 5, 2019)

**Abstract**

An extension of the Bachelardian notion of epistemological profile allows us to establish the appearance of a strong 'conformist thinking' component following certain formulations of the Galilean concept of the relativity of motion, namely those associated to the idea that the Sun might revolve around Earth from certain points of view. We deduce the existence of a strong remanent paradigmatic pressure related to the Copernican paradigm. A study involving more than 2900 students, primary school teachers and future teachers shows a 97% prevalence of (pre-Galilean) Copernican conformism. The epistemological profiles of several high level academics, including professional physicists, a mathematician and a sociologist show that, under strong paradigmatic pressure, even professional scientists can shift from a scientific to a conformist way of thinking, possibly related in some cases to a 'gate-keeper' posture. Even when desired, the personal shift to a more advanced (relativist) understanding of such a high-pressure issue as heliocentrism can be a lengthy process. Some consequences of this disturbing cognitive process for science education are also discussed.

# I. INTRODUCTION

Although scientifically closed for four centuries, the question of heliocentrism still puzzles many a layman. It has now become a common journalistic device to trick people about their scientific literacy: Bringing someone to admit that he still believes that the Sun might revolve around the Earth is a classic joke even in mass media. An eloquent example is an excerpt from the French edition of the TV show *Who Wants to Be a Millionaire*, in which, to the question '*What gravitates around the Earth?*', a candidate answers '*The Sun*', after a majority vote of the public in this direction. His unfortunate answer instantly became viral worldwide. To this date, it has triggered almost 600 contemptuous comments on the illiteracy of the candidate (on nearly 220,000 views for a single Youtube rendering [1]). More generally, polls regularly conclude that a significant minority of the population (typically 20 or 25% [2, 3]) remain pre-Copernican.

Or do they really? Beyond the cheap mockery, or perhaps because of it, everything happens as if the Copernican quarrel had left deep enough traces in the collective scientific imagination to make such answers difficult to interpret. How strongly influenced by the social and cultural context are they? Would they remain the same for different, less triggering formulations of the same physical questions, or in slightly different contexts?

While social influence is known to be '*a pervasive force in human social interaction*', yielding individuals to '*modify their opinions, attitudes, beliefs, or behavior towards resembling more those of others they interact with*' [4], even when purely perceptual judgments are involved [5], or even when there are no other member of a given social group nearby [6], and has been the focus of many studies in the field of psychology [6, 7], it is seldom considered in scientific matters.

On the contrary, scientifically minded people would often claim to be able to abstract themselves from such parasitic mundanities in order to deal objectively with scientific matters, a claim which may seem generally confirmed by the scientific community's recognized ability to reach rapid consensus once all the facts are known about a given question [8, 9].

Yet, the example of heliocentric thinking might endanger this comfortable conclusion since it suggests that, in some cases, even well informed individuals might occasionally participate to an otherwise obsolete paradigm.

The purpose of this article is to determine to what extend people may defend notoriously scientifically obsolete positions, whether or not they're familiar with the current ones ; what might trigger such epistemological skidding; how deeply and in what proportions people might be affected; and whether a solid scientific background and/or daily practice are sufficient to prevent them from reverting to some kind of non-scientific thinking. It also considers some consequences of this disturbing cognitive process in science education.

## A. Paradigmatic pressure

Besides controlled scientific thinking and established scientific knowledge, all kind of parameters — sociological, psychological, linguistic, political, ideological, organisational, and so on — seem susceptible to affect the way a person thinks about scientific issues. It is customary, when such supposedly parasitic parameters are even considered, to assume that their total influence is generally indifferent, either because they are intrinsically negligible or because their influence randomly cancel each other out [10].

Since we can not identify and describe them in detail, we propose to group their cumulative influence under the umbrella expression of *paradigmatic pressure*. Qualitatively, the pressure imposed by a given scientific paradigm will be considered weak if a reference (direct of indirect) to a key issue of this paradigm doesn't affect the average response to a pertaining scientific question; or strong if it induces a significant change (assuming a fully solved question, with a duly established and consensual answer among the scientific community).

Note that while the notion of paradigmatic pressure strongly overlaps with the social-cognitive notions of *conformity* (i.e. the act of changing one's behavior to match the responses of others, often in order to '*enhance, protect or repair [his] self-esteem*'), and *compliance* (i.e. acquiescence to a request) [7], it is not limited to social interactions, but may also derive from one's own knowledge, scholarly or not. Moreover, by analogy with other kinds of group pressure [5], different personalities, either more 'yielding' or more 'independent', may *a priori* present very different individual responses to the same paradigmatic pressure.

## B. Copernican vs. Galilean paradigm

As noted, the Copernican paradigm appears at first glance as one of the most likely to enforce a particularly strong paradigmatic pressure, yielding to an '*extreme consensus*' and strong group polarization [4], due both to its mediatic status as a scientific literacy litmus test, to its historical position at the dawn of modern science both as the archetype of a scientific revolution [11] and as the stake of a critical quarrel between science and religion, and generally due to its deep embedment in our present popular culture.

We will thus focus our study on the reception of scientific propositions more or less explicitly related to Copernican heliocentrism (1543 [12]), as opposed to the more modern relativist approach, as first formulated by Galileo: '*For all things that also participate to a motion, this motion doesn't act, it is a if it was not.*' (1632 [13]) While the former approach discuss the situation of the 'center of the world' (still an Aristotelician notion), shifting it from the center of the Earth towards the center of the Sun, the latter acknowledges that both positions and in fact every reference frames, are equally legitimate, depending on the *point of view* of the observer. While current time physics would generalize this 'Principle of Relativity' [14] in a more abstract 'general covariance principle', it can still be popularized as: '*There is no physically privileged state of motion*' [15].

Anyway, while '*the concept that the Earth revolves around the sun [...] was once an exciting and disturbing possibility that inspired new actions and transformed the way people thought about the Earth and humanity in some surprising ways*', it nowadays seems '*so ordinary that we do not even think about it twice*' [16]. It has even been suggested that '*the position of the Copernican system during the scientific revolution was that of an icon, a rallying-cry, a point of reference, or even a shibboleth. The various arguments for or against proved less important than its simple presence as an ideal, a challenge to traditional learning or a symbol of the new.*' [17] and that '*The great Copernican cliché*' might even, nowadays, be more metaphysical than physical in nature, being '*premised upon an uncritical equation of geocentrism with anthropocentrism*' [18].

It is interesting to note that while the notion of a fixed Sun might, from an epistemological point of view, seem to be in contradiction with the common daily experience of seeing the Sun move in Earth's sky, and thus to the primacy given to observation by empiricists, more rationalist minds may, on the contrary, consider the supposedly 'objective' (in a Newtonian perspective, also obsolete from an Einsteinian one) rest of the Sun to be better founded than the 'subjective' personal experience of its apparent motion [19].

If its purely scientific features nevertheless remain central in the more controlled context of science education, the diagnostic remain quite alarming. It has been shown that pre-Galilean (mis)conceptions — namely, the existence of a privileged reference frame, whether heliocentric or geocentric — are even more widely shared, and not only that '*many students hold alternative notions in various basic astronomy subjects*' [20], but also that many teachers and prospective teachers of physics '*show a deep lack of understanding of both concepts of reference frame and event*' [21], and have limited understanding of the Earth-Moon-Sun scale and geometric configurations [22]. The Copernican model remains seen as '*the final essence of astronomical concepts*' [23]. Saltiel [24] established that for most students, '*proper motion and immobility are defined intrinsically, and not with respect to specified bodies and frames*' and that '*motion and rest are thus fundamentally inequivalent, a typical pre-Galilean view.*' Further difficulties relative to the Newtonian distinction between 'real' and 'apparent' motion are also well documented [25–27]. Panse, Ramadas and Kumar also established that '*the more prevalent alternative conceptions are also the ones which are held with greater conviction*' [25, 28].

Concerning the principle of relativity itself, it has also been shown that even physics students enrolled in a course on special relativity tend to trivialize it by reducing it to a strictly kinematic interpretation [29].

## II. A CONTEXT-DEPENDENT RELATIVISM

### *A. Methodology*

In order to establish a baseline, a questionnaire was submitted to a sample of 2115 French first year college students in science (either universitary L1 or from the more selective *mathématiques supérieures*). The questionnaire included (among 102 other questions about science, maths and other skills) three closely related formulations about relativity in motion (Table I), to be validated or not. It was organized over 4 years through a running call from the French Physical Society (*Société Française de Physique*, SFP) to volunteer teachers.

Respondents were informed at the beginning of the course or training that their answers would remain anonymous and not be rated, but used only for statistical purposes [30].

The first question dealt with the problem *in abstracto*; the second one involved a child on a merry-go-round, reminiscent of typical classroom physics problems and presumably eliciting only a weak paradigmatic pressure; finally, the third formulation directly involved the relative motion of the Earth and the Sun, and was thus likely to bring out the full Copernican paradigmatic pressure, if any, by focusing the attention on the salient Copernican norm and thus making it focal [6]. The latter (Form. 3) was presented as an open question, while the first two (Form. 1 & 2) were multiple choice questions (MCQ): *Yes / No / I don't know*.

| Formulation 1, MCQ (abstract) | *The description of the motion of an object depends on the observer who describes it.* |
|---|---|
| Formulation 2, MCQ (presumably weak PP) | *A child is on a merry-go-round. Sitting on a bench, his mother sees him rotating at a constant speed. From the child's point of view, it is not the merry-go-round but his mother who turns around him.* |
| Formulation 3, open question (presumably strong PP) | *A person observes the motions during the day of the Sun in the sky and of the shadow of a stick on the ground. He argues that the Sun turns around the Earth in 24 hours. What do you answer?* |

TABLE I. Three closely related formulations about relativity in motion, eliciting various levels or paradigmatic pressure (PP)

For each tested formulation, answers were distributed among two broad classes labeled either 'Galilean' or 'Copernican', depending on whether they validated or not the proposed formulation. Note that most answers were very brief, and that these labels do not imply a direct reference to either Nicolaus Copernicus or Galileo Galilei, to their work or to their philosophies. A third broad class, 'Other', was added when more ambiguous or irrelevant — including *'I don't know'* —, but definitely non-Galilean, answers were elicited.

For comparison, either Formulation 3 or an alternative Formulation 4 (but neither 1 nor 2) was also submitted, also through a questionnaire, to various other populations, including 357 French primary school teachers and 269 other students, from high-school (21) to post-graduate school (134). Note that while Formulation 3 directly referred to personal observations, Formulation 4 rather emphasized the possibility of diverging opinions. (Table II). Finally, we submitted to an additional population of 166 post-graduate students in a primary school teaching track ('M1 & M2 MEEF') a comprehensive formulation directly referring explicitly both to a personal experience and to the notion of point of view (Form. 5), for a grand total of 2907 participants.

| Form. 4a, open question (presumably strong PP) | *Two people are talking. One argues that the Earth revolves around the Sun, the other that the Sun revolves around the Earth. They turn to you for your opinion. What do you answer them?* |
|---|---|
| Form. 4b (*idem*) | *Two pupils are talking. One argues that... [same as 4a]* |
| Form. 5 (*idem*) | *Two pupils argue: the first states that the Earth does revolve around the Sun and that it is not possible to say that the Sun revolves around the Earth. The second one insists that, from a Terran point of view, we have the right to say that the Sun revolves around the Earth; he adds that when we look at the shadow of a stick planted in the ground during we can see his shadow moving throughout the day. Which pupil do you agree most with?* |

TABLE II. Same as Table I, Additional formulations

## B. Results

### 1. SFP Questionnaire

A first remarkable result is that *none* of the 2115 students involved in the SFP study chosed the '*I don't know*' answer proposed on the multiple choice questions.

The first, more abstract formulation induced a strong majority (88%) of correct (Galilean) answers (Table III). This is an expected result, or even a somewhat disappointing one for science college students since, according to French programmes [31], the skill '*Understand that the nature of the observed movement depends on the chosen frame of reference*' is in fact expected as soon as the first year of high school, and work on it is often continued during the two additional years from there to college.

| Population | Form. | Nb | Copernican | Galilean | Don't know |
|---|---|---|---|---|---|
| 1st year college science students | 1 | 2115 | 11.9% | 88.1% | 0 |

TABLE III. Proportion of students validating ('Galilean' answer) or not ('Copernican') the proposition that the description of a motion depends on the observer.

The concrete application of this principle of relativity nevertheless already appears a bit more difficult, even within a presumably weak-pressure context, since 14% less student (74%) manage to also validate Formulation 2 (Table IV).

| Population | Form. | Nb | Copernican | Galilean | Don't know |
|---|---|---|---|---|---|
| 1st year college science students | 2 | 2115 | 26% | 74% | 0 |

TABLE IV. Same as Table III, but with Formulation 2
(merry-go-round, weak paradigmatic pressure)

While they confirm the seemingly pessimistic estimations of the medias about scientific literacy [2, 3] (even if our questionnaire oriented incorrect 'non-Galilean' answers towards Copernican rather than towards pre-Copernican answers), these first results must be immediately relativized.

Not only does the more abstract Formulation 1, closer to those offered by manuals, diminish the proportion of wrong answers by more than a factor 2 with regards to a more concrete formulation (Form. 2), but even those among the first year college science students who gave `informed' Galilean answers regarding both formulations 1 & 2 (1424, or 67.3 % of the initial 2115) will massively shift to a 'Copernican' stance when confronted to a proposition involving the relative motion of the Earth and the Sun (Form. 3), with less than 4% maintaining a Galilean stance in this modified context, as shown in Table V.

| Population | Form. | Nb | Copernican | Other | Galilean |
|---|---|---|---|---|---|
| Previously 'Galilean' | 3 | 1424 | 76.4% | 19.8% | 3.8% |
| Previously 'non Galilean' | 3 | 691 | 69.8% | 28.9% | 1.3% |
| *Total* | *3* | *2115* | *74.2%* | *22.8%* | *3.0%* |

TABLE V. Repartition of 1st year college science students
answers to Form. 3, depending on their answer to the MCQ

Note that, while it appeared that all answers to Formulations 1 & 2 spontaneously distributed themselves among the two main classes of the MCQ, it was not the case with the open question (Form. 3), which yielded a significant proportion (22.8 %) of more ambiguous responses, hereafter labelled 'Other', among students having previously validated both Form. 1 & Form. 2 ('previously Galilean') as well as among those who had not ('previously non Galilean'). In some cases, this ambiguity might be semi-deliberate and derive from a reluctance to give either a clearly 'Copernican' answer in obvious contradiction with the previous two, or a clearly 'Galilean' answer in obvious contradiction with the dominant Copernican paradigm.

Anyway, these results unambiguously validate our hypothesis that the answer to a given scientific question strongly depends on its formulation. The interpretation of the answers to such questions thus calls for a very cautious analysis of the context in which they are asked.

But the key result of this first study is that an overwhelming 97% of this large sample of college-level students appear unable to apply correctly the same idea they were more than 88% to validate earlier in the same questionnaire, as soon as the Earth-Sun problem is called on by Formulation 3, which was deliberately constructed to see if the participants would find themselves able validate the brutal proposition that '*the Sun turns around the Earth*', hopefully by introducing either the (implicit) notion of frame of reference, or the notions of observer or of point of view (explicit in Formulations 1 & 2, respectively); or if they would be so disturbed that they'd fall back into Copernican thinking.

## 2. Other populations

We obtained similar results for 792 additional participants from various students and teacher populations, including 21 first year high school students and 114 second and third year college students following an optional course in science education, 300 post-graduate students in either a primary school or a high school physics teaching tracks, and 357 in-service primary school teachers (Table VI).

Little more than 3% of the total population surveyed (students of various levels and teachers, Table VI) use the Galilean concept of relativity of the movement to answer the question. Although students learning to teach physics perform slightly better than other participants, nine out of ten of them remain Copernican.

The physics teacher of the 21 high school students questioned was surprised, and disappointed, that only one of them mobilizes the principle of relativity, although she had them working on it the previous week, and on an astronomical example. It is interesting to note, however, that while the work done in class had not been met with particular reservations, she had chosen to apply it to the description of the relative motion of Venus and the Sun, a problem mathematically equivalent to the relative motion of Earth and the Sun but which, unlike the questionnaire, does not constitute a direct reference to the Copernican problem.

| Population | Form. | Nb | Copernican | Other | Galilean |
|---|---|---|---|---|---|
| 1st year high school students | 4 | 21 | 90.4% | 4.8% | 4.8% |
| 2d & 3rd year college students | 4 | 114 | 98.2% | 0 | 1.8% |
| Students in a primary school teaching track | 4 | 103 | 83.5% | 16.5% | 0 |
| Students in a primary school teaching track | 5 | 166 | 89.8% | 0 | 10.2% |
| Students in a HS physics teaching track | 4 | 31 | 90.3% | 0 | 9.7% |
| Primary school teachers | 3 | 357 | 91.6% | 5.6% | 2.8% |
| *Total Other populations* | | 792 | *91.0%* | *4.8%* | *4.2%* |
| *Total simple (including SFP)* | | 2907 | *78.8%* | *17.9%* | *3.3%* |

TABLE VI. Same as Table III, but with a question about
the relative motion of the Earth and the Sun (Form. 3, 4, or 5)
and for various student and teacher populations

While the more explicit and yielding Formulation 5 appears to limit the proportion of ambiguous 'other' answers and to induce the highest proportion of 'Galilean' answers among future primary school teachers (Table VI), the latter nevertheless remain limited around 10 %.

**C. Discussion**

At any rate, this first part of the study confirms a massive dependence of the participants answers to a given physical problem on the precise formulation of the question.

Even if one can not exclude, at this level, some contribution of a simple misunderstanding of the principle of relativity in motion, it also confirms that, as expected, scientific judgment can be strongly affected by the simple mention of a possible motion of the Sun around the Earth, regardless of the fact that this is a fully legitimate point of view within the Galilean relativist paradigm (at least as soon as a Terran point of view is specified), although in total contradiction with the older Copernican paradigm. The understanding and appropriation of the concept of relativity by the participants to the study thus appears highly context-dependent. Considering the size and diversity of the sample, it seems that this result can safely enough be extended to the general (i.e. non-specialized) public.

An immediate corollary, also expected considering the historical weight of the Copernican quarrel, is that the pressure imposed by the Copernican paradigm is very strong and significantly higher than the pressure associated to the relativist paradigm *per se*.

However, it also induces a very strong (and mostly correct) perception of Copernican views as highly consensual, which makes it difficult to distinguish between the participants' conformist thinking, their likelihood of expressing a prejudice being strongly reinforced by their perception that most others would approve their response [32], and their compliance to the answer they believe the (presumably teacher) author of the questionnaire might expect [7].

It is interesting to note that while Formulation 3 underlines direct personal observations (of the motion of the Sun in the sky and of the shadow of a stick on the ground), and might thus have its legitimacy reinforced by the principle of primacy of observation, it doesn't induce more Galilean answers than Formulation 4, which doesn't. Paradigmatic pressure thus appears able to distort judgment regardless of personal experience, like group pressure in the classic Asch's conformity experiments [5].

### III. EPISTEMOLOGICAL PROFILES

#### A. Diversity of pathways to knowledge

The scientific method is historically, at least in Western cultures, the third major access path to knowledge, after the immediate and the scolastic paths. We call here 'immediate' any direct access, non mediated, to knowledge, whether it is an *a priori* form of the latter, according to transcendental æsthetics [33]; a mystical evidence, in the religious acception of the term; a perceptive *a posteriori* judgement ('*The sky is blue*'), or even a learned skill or a social norm (i.e. '*certain standards, expectations, and rules for what is "normal" or "appropriate" to feel, think and do*' in a given social group [6]) embedded enough to have almost become a reflex. The scholastic path explicitly lies on some external authority (typically, in the Middle Ages, Aristotle's teachings), or more generally of some external

*Blanquet & Picholle — Paradigmatic Pressure — PRPER* 11

discourse. The scientific path, on the other hand, lies on the refusal of any argument of authority and on the rational examination of any relevant argument.

It is important to note that these three pathways are in practice likely to be associated, in varying proportions, for a given individual and about a given question. Besides, even an hypothetical ideal scientist may justify his private religious faith by an epiphany, immediate in nature, and his personal political convictions by a scholastic adherence to a body of doctrine. Even better, as far as its very scientific activity itself is concerned, his epistemological profile may vary depending on the considered concept, from empiricism to discursive rationalism [34, 35].

We also know that these proportions, and more specifically the relation to the authority of already established knowledge, can strongly depend on the context: the same ideal scientist will not generally have the same spontaneous epistemological relationship to a knowledge new to him, depending on whether it is a canonical cultural object, duly validated by the corresponding scholarly culture ('World 3', in Karl Popper's lexicon [36]); an unprecedented or unconfirmed research result, to which he will oppose a resolute methodological skepticism; a heuristic speculation ('*Anything goes!*' [37]); or an actual revolutionary claim (in the sense of Kuhn [38]), breaking with the dominant paradigm.

Other more cunjunctural parameters can also affect the response of a scientist to a given proposition. For instance, in the United States, the question of 'intelligent design' has exacerbated an internal tension between science and religion, to profoundly modify the scientific discourse, and teaching postures, on these sensitive questions [39]. On a different note, some more quirky questions, such as those of perpetual motion or astrology, may induce more sociological than properly scientific 'gatekeepers' postures among professional scientists [40, 41]).

### B. A Bachelardian approach

#### *1. Identification of the epistemological components*

It is indeed to account for this variability of the epistemological posture of a same individual with regard to different concepts that Gaston Bachelard develops the notion of epistemological profile in his *Philosophy of the No* [34]. He observes that we retain traces of the steps we have taken to arrive at the understanding of a concept. He thus identifies the crossing of successive obstacles to arrive at our current understanding of the idea of mass and associates these difficulties with the progressive overcoming of successive philosophical levels — naïve realism, clear empiricism and positivism, classical rationalism, complete rationalism — to achieve what he calls 'discursive rationalism' (or 'surrationalism'). He then proposes to give an account, for a given individual and a given scientific concept, of these different philosophical levels in his appropriation of the notion by an 'epistemological profile'. Established in a partially subjective way, it includes '*on the x-axis the successive philosophies and on the ordinate a value which — if it could be exact — would measure the frequency of use of the notion, the relative importance of our convictions*' [34].

- For Bachelard, a symptom of a *naive realistic* relationship to knowledge is '*the speed with which [a concept] is understood*' and mobilized under '*imprecise conceptual forms*' [34]. Thus, a subject will be satisfied to describe a movement 'in absolute', without precision of the reference frame of study or by using a 'natural model' [24]. In the same way he will be able to content himself with explaining the 'broken' appearance of a stick immersed in water by the existence of two media of different optical index.

- *Clear and positivist empiricism* translates according to him into a simple 'conduct' that is a '*necessary and sufficient reference*', such as the search for all forces present in a given frame of reference to describe a motion, or, in optics, a relation to the concept based on geometric optics experiments (typically in terms of optical index) or on the historical experiments of Fizeau, Morley & Michelson, etc.

- *Classical rationalism* is characterized by the rational integration of a concept into a '*body of notions*'; '*It's the time of notional solidarity*' in Newtonian mechanics or Maxwell's theory of electromagnetism.

- *Complete rationalism* proceeds through '*an opening into the interior of the notion*' which can then be analyzed; '*It is touched by relativity: an organization is rational relative to a body of ideas*'. It is the time of special relativity or the taking into account of complex propagation regimes of light (nonlinear optics, disordered or metamaterial media, etc.).

- *Discursive rationalism*, the ultimate opening of the dialectical philosophy of *why not?* does not refrain from questioning the very foundations of concepts, and, here, takes into account arguments pertaining for instance to general relativity (e.g. Mach's bucket thought experiment [42]) or quantum field theory.

In the context of the present study, two additional components should be considered, namely:

- *Conformist thinking*, or automatic commitment to what one understands would comply to a given paradigm. Note that conformist thinking, of scholastic essence, differs from classical rationalism only in that the concept is integrated in the paradigm's body of notions without conscious rationale, or only through irrelevant, emotion-driven pseudo-reasoning.

- and *Immediate thinking*, informed or not.

The components originally proposed by Bachelard made it possible to identify a rather comprehensive structure of the 'scientific' posture of an individual with regard to equally scientific concepts. Strictly speaking, it would probably also be appropriate to dene more precisely, with the aid of several adapted subdivisions, the postures 'immediate' and 'conformist'. However, our purpose being limited to a first identification of the influence of the paradigmatic pressure, we will stick to this gross grain and limit here the field of application of this tool to professional scientists whose informed immediate thinking, for instance, can not be confused with a lack of ability to justify an answer.

*2. Original Bachelardian tool*

As an illustration, the philosopher established his own epistemological profiles relating to the concepts of mass and energy and shows that they present differences related both to his culture and to his personal and daily experience of the two concepts (Figs. 1 & 2).

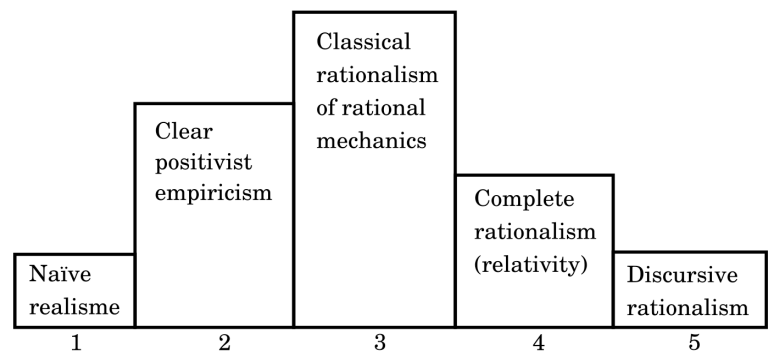

FIG. 1. Gaston Bachelard's own personal epistemological profiles relative to the concept of mass, as determined by himself in [34] (*fac simile*)

The French philosopher of science considers himself mostly classically rationalist about both notions. For him, '*to the extend that it is a clear notion, the notion of mass is above all a rational notion, which was shaped during a period of classical education and developed during a long period of teaching elementary physics*', while he '*gives a greater importance to the dialectized concept of energy [since it] has found its realization already, which is not the case for the concept of mass*'; however, '*as if to compensate, there subsists in us a confused knowledge or energy [...] made up of a mixture of pig-headedness and of rage [...]*'.

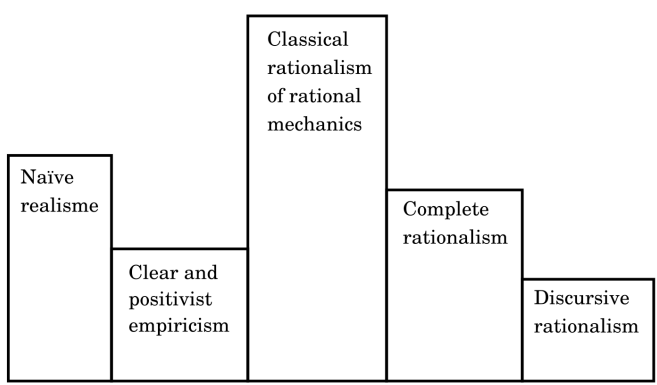

FIG. 2. Same as Fig. 1, but for the concept of energy

It should be noted, however, that if epistemological profiles allow, to a certain extent, a rough quantitative approach to the paradigmatic pressure, then it will only have the same value as in regard to a given subject and about a given concept.

*3. Expanded profiles including conformist thinking*

The primary object of *The Philosophy of the No* is to discuss how enlightened minds, endowed with a solid scientific culture and a ne understanding of the nature of science, are likely to react to the emergence of new concepts (non-Lavoisian chemistry, non-Euclidean geometry, etc.). Since Bachelard himself encouraged his readers to '*use philosophical elements detached from the systems in which they originated*' [34], a first generalization can be to adapt the approach to less sophisticated audiences, such as teachers primary education, of whom it is possible to establish the epistemological profiles - most of them empiricist - but at the cost of a considerable sweetening of the philosophical 'levels' considered [35].

In the same spirit, insofar as we are interested here in the relation to authority and the possible interference of non-scientific access routes to knowledge, we will allow ourselves to add two components to the Bachelardian schema: evaluating the relative importance of a conformist thinking in the relation of the subject to a given concept ; and of an immediate thinking component. While we place them for simplicity at the left of the diagram, this does not imply any anteriority in the subject's personal conceptual history. The relative weights of the epistemological components is still relative (hence arbitrary units).

By construction, this addition does not change the epistemological profiles of the philosopher, whose entire works show that he never gives in to these modalities; indeed, their weight is therefore essentially null for him with regards to the concepts considered (Fig. 3).

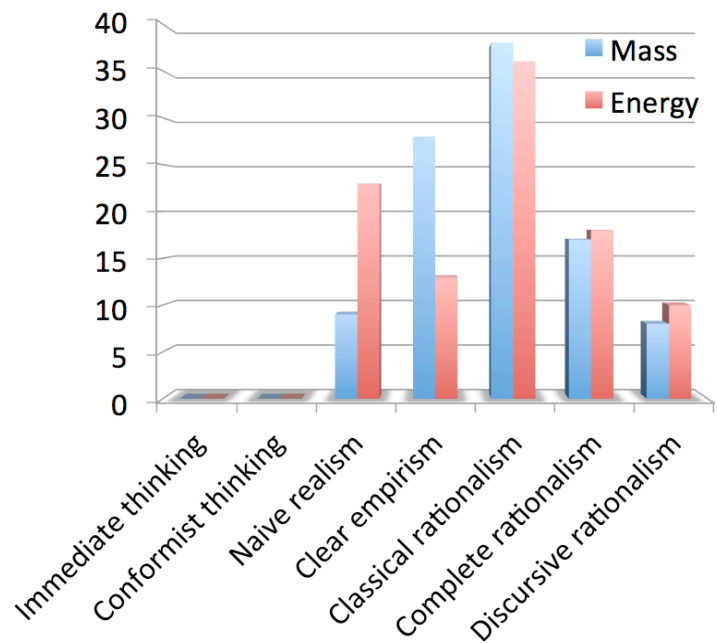

FIG. 3. Gaston Bachelard's expanded personal epistemological profiles relative to the concepts of mass (left, blue) and energy (right, red). Same as Figs.1 & 2, respectively, but with two additional (null) components to the left (arbitrary units)

## IV. METHODOLOGY

### A. Identification of 'sensitive' concepts and formulations

We will consider more specifically two well-established physical concepts, fully consensual in the scientific community. To test the importance of the formulations and the cultural context, each one will be approached from two different angles, chosen according to the cultural autonomy that they may have acquired, or not, beyond their initial scientific framework.

The first of these concepts is the velocity of light. Object of polemics in the seventeenth century (does light spreads more or less quickly in a material medium than in a vacuum?), it stabilized rapidly in the eighteenth where, with the maturity of the Geometric Optics, it was reinterpreted in terms of optical index, now defined as the ratio between the velocity of light in a vacuum ($c$) and the phase velocity of a wave in a material, $n = c/v_\phi$. Then, with the reinterpretation of $c$ as the limit speed for the transport of energy and information within the framework of the special theory of relativity (1905), whose pseudo-paradoxes fascinate the general public, the need arises to distinguish several definitions of the speed of light (group, signal, front velocities, etc., fully understood since 1914 [43]).

We will therefore try to compare, for the same subjects, epistemological profiles relating to the same concept of velocity of light, first considered in the presumed neutral formulation of Geometric Optics (i.e. in terms of optical index), then in the possibly more disturbing formulation of 'velocity of light', which 'everyone knows' cannot be exceeded [44] — even though various recent experiments have demonstrated superluminal group and signal velocities [45].

The second concept is that of relativity of motion. At the heart of the Galilean revolution, in its astronomical formulation, it debunks the Copernican quarrel and opposes the deep cultural anchoring of the idea that 'the Earth revolves around the Sun' (and *not* the other way around!). We still find signatures of this way of thinking in the mocking reactions faced by people who ignore them. Reformulated in more neutral terms, widely used in the usual courses and manuals of mechanics (scientific terminology, first year of college) in the context of Newtonian physics, it typically solves problems concerning the relative movement of trains or of rides and involves the choice of a reference frame for the description of a motion and/or the comparison of the description of a motion in different reference frames.

Here again, we will compare epistemological profiles relative to the same concept of relativity of motion, considered first in the presumably neutral formulation of classical mechanics, then in the most disturbing, even anti-Copernican, of the question of the movement of the sun around the Earth.

### B. *A priori* estimation of the associated paradigmatic pressure

An attempt to qualitatively identify the frequency of references to these concepts makes it possible to advance a first hypothesis on the different relative paradigmatic pressure levels:

- the concept of optical index, which is rather technical, is not associated with a specific paradigm (it is used as well in geometrical optics as in the theory of Maxwell's electromagnetism, and even in quantum optics). It does not seem to be particularly mediatized and one can expect a low paradigmatic pressure. In terms of conformist thinking, we can associate limiting prejudices such as '*The optical index is always positive and greater than 1*' (actually an invalid proposition in some exotic metamaterials [46]);

- the concept of velocity of light is today associated with the Einsteinian paradigm. Beyond the scientific community, references to the Einsteinian paradigm remain conned to certain audiences (such as 'geeks' and science fiction fans) and specialized programs, and moderate paradigmatic pressure can be expected. Associated formulation: '*One can never exceed the velocity of light*';

- the principle of relativity, considered from the mechanistic point of view, is associated with the Galilean paradigm. This problem is essentially absent from the media. We can therefore expect a weak paradigmatic pressure. Related conformist wording: '*All physical reference frames are equally legitimate*'.

- the same, considered from the astronomical point of view, remains largely associated with the Copernican paradigm (although Galileo's name is often associated with it: '*And yet it moves!*'). Associated formulation: '*The Earth revolves around the Sun* (implicitly: *and not the other way around*)'. References are frequent and easily provoke passionate responses: this suggests a strong paradigmatic pressure.

### C. Description of the sample

To test these working hypotheses, we have chosen to construct the expanded epistemological profiles of a sample of high-level academics and intellectuals, including two experienced physicists to avoid any bias related to possible problem comprehension concepts. The establishment of broadened epistemological profiles is based on interviews, e-mail exchanges and personal knowledge of the participants in the study.

These epistemological profiles were co-constructed by the two authors in a concerted manner. For each subject, the importance of each component was evaluated independently and then the ordinates were normalized (sum of the different components = 1, arbitrary scale on the ordinate axis but identical for all the profiles) to facilitate the comparison of the profiles.

Our sample consists in four high-level academics (Table VII), presumably less likely than students to conform to other people's opinions, since their own professional achievements should constitute a sufficient source of self-esteem [47]:

| Prof. A | Prof. B | Prof. C | Dr. D |
|---|---|---|---|
| Physicist | Physicist | Mathematician | Sociologist |

TABLE VII. Respective Fields of the four Academics

First, we interviewed two senior physicists, university professors emeritus and successive Heads of the same large research laboratory, in which the interviews were conducted. Since one of the authors (EP) is also a physicist, the interviews were based on mutual professional understanding and respect. Both Maxwell's theory of electromagnetism and Einstein's special theory of relativity are familiar to them, as are experiments based on the principle of a limit speed of light.

Professor A has been Head of the Physics Department of his university. An experimental physicist, his problem-solving approach often goes through an hands on representation / visualization. Generally recognized for his pedagogical skills, he devoted considerable effort to the improvement of the understanding of mechanics by his students.

Prof. B has also taught mechanics and optics to undergraduate university students. She is the author of popular science books. An experimental physicist with a strong theoretical background, writing down equations is nevertheless often her first reflex to solve a scientific problem.

The other two members of our sample were interviewed during an interdisciplinary workshop combining science and fiction.

A university professor in mathematics, Prof. C is deputy head of a major Maths laboratory. He is passionate about science fiction and used to thinking 'outside the box'.

A sociologist, Dr. D is research director at the French CNRS and deputy head of a large Sociology laboratory. Dr. D is very interested in the sociology of science and of education. A former science undergraduate, he claims to remember his physics classes.

# V. SOME EFFECTS OF THE STRONG COPERNICAN PARADIGMATIC PRESSURE

## A. Effect on two physicists

### 1. Expanded epistemological profiles on the concept of velocity of light

Figures (4) and (5) show the expanded profiles of the Profs. A and B relative to the concept of velocity of light formulated respectively in terms of index and propagation. A first observation is the similarity of these profiles, dominated by the classic clear and positivist empiricism and rationalist components.

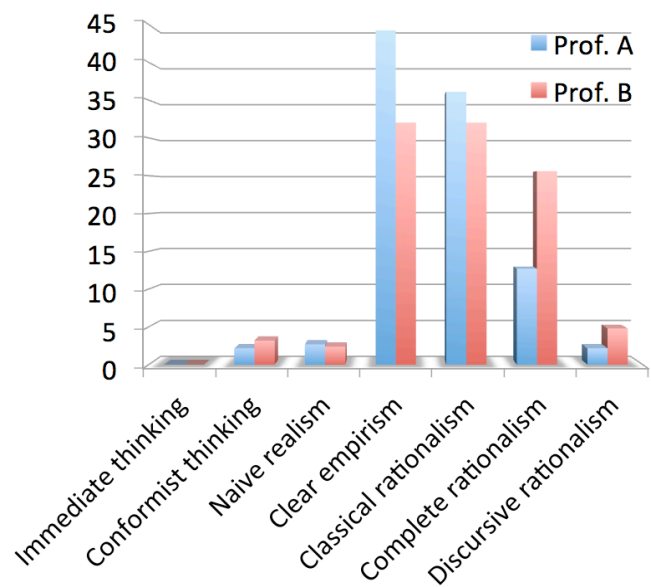

FIG. 4. Expanded epistemological profiles of Profs. A (left, blue) and B (right, red) relative to the concept of velocity of light formulated in terms of index (arbitrary units)

Passionate about the history of science, Professor A was interested in the history of optics and the historical experiences he knows in depth. We found in him an empiricist component more marked than in Professor B, who tends to favor formal reasoning.

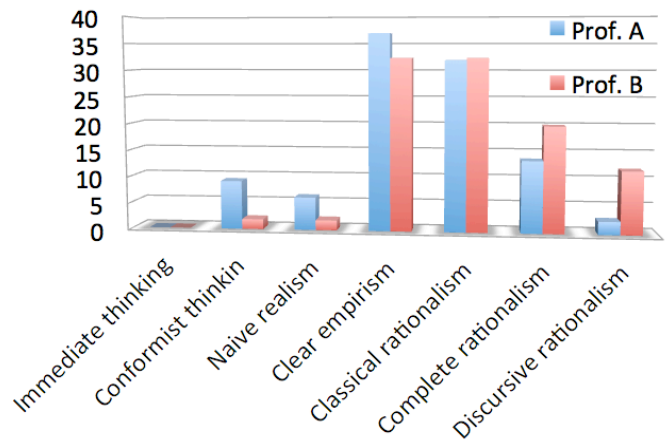

FIG. 5. Same as Fig.4, but now formulated in terms of (superluminal) propagation

Both present a significant share of complete rationalism on the concept of propagation of light, possibly related to the fact that, as part of their management duties, they became aware of work on the superluminal propagation of light pulses carried out in their laboratory.

While the 'discursive rationalism' and 'conformist thinking' components remain rather low in both profiles when approaching the concept in terms of index, the presumably moderate paradigmatic pressure associated to the issue of superluminal propagation nevertheless appears sufficient to yield a moderate, yet significant on the profile of Prof. A, who visibly appears uncomfortable when one speaks of phenomena of superluminal propagation, even if he fully understands and readily admits the mechanisms at work. Namely, while the general allure of his profile remains stable, a noticeable conformist component appears, mostly at the expense of the empiricist and classical rationalist components (Fig. 6).

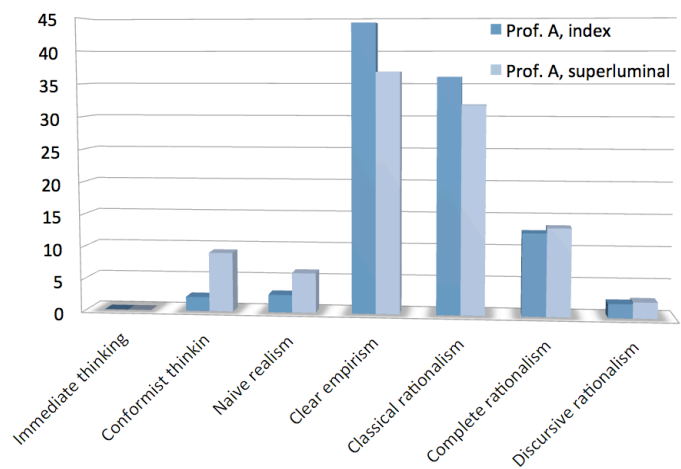

FIG. 6. Emergence of a conformist component under moderate paradigmatic pressure: Prof. A's epistemological profiles relative to the concept of velocity of light formulated in terms of optical index (low PP, darker, left) or superluminal propagation (moderate PP, lighter, right)

Prof. B's epistemological profile remains mostly stable with both formulations.

*2. Concept of the relativity of motion considered from a mechanical point of view*

The profiles of the same Profs. A and B are now being considered when they are asked to mobilize their knowledge on issues related to the relativity of movement, such as the description of that of a person on a carousel in different frames of reference or that of the relative movement of the Earth and the Sun.

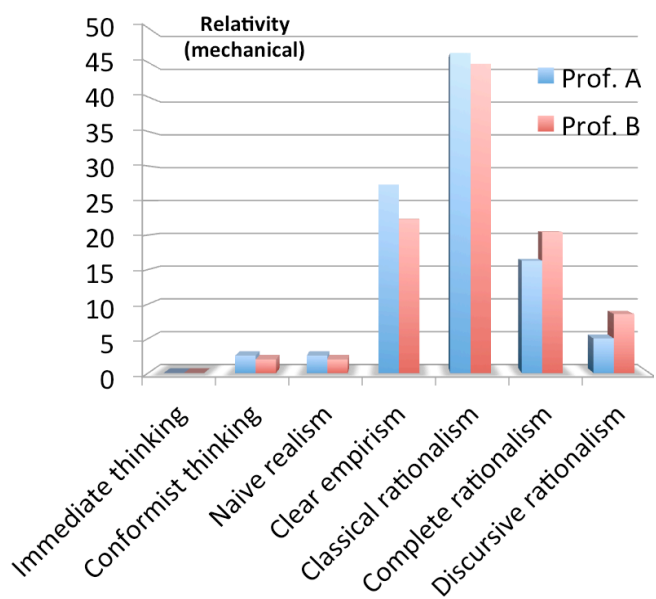

FIG. 7. Expanded epistemological profiles of Profs. A (left, blue) & B (right, red) relative to the concept of relativity of motion formulated from a mechanical point of view (arbitrary units)

Both profiles relative to the concept of relativity of motion formulated from a mechanical point of view (Fig. 7) are again rather similar and present a classic rationalist dominant associated with a strong part of clear and positivist empiricism perfectly adapted to the formulation of the strictly 'classical' problem proposed. They correspond to the intuition a priori of the profile of academics based on the experience of a long career teaching Newtonian mechanics, with which they both are perfectly at ease; and maybe especially of French academics, since '*the substitution of mechanics or mathematical physics for theoretical physics*' has sometimes been considered '*a salient feature*' of the French reception of relativity [48].

They also present comparable rationalist components, both 'complete' and 'discursive'. None of them conducted research in this area of Einsteinian relativity, but both have a broad knowledge of modern physics. Prof. B, more accustomed to theoretical discourses, seems to have more confidence than Prof. A in her own understanding of the theories of special and especially general relativity, the latter taking the time of reflection, punctuated by several '*Wait, wait, wait!*'

Finally, the residual part of conformist thought related to the invocation of the authority of Einstein or Galileo during an exchange on the relativity of the movement accounts for the acquisition of consciously assumed automatisms of thought, sometimes used as shortcuts.

*3. Concept of the relativity of movement considered from an astronomical point of view*

The epistemological profiles of the two physicists with regard to the relativity of the movement of the Earth and the Sun, considered from the very disturbing point of view of the geocentric reference (Fig. 8) appear on the other hand very different, between them as with regard to the profiles previously obtained from a mechanic point of view (Fig. 7). If the empiricist and rationalist classical components remain dominant in the Prof. A and important in the Prof. B, the complete and discursive rationalisms practically disappear, whereas a very important conformist thinking component appears in both.

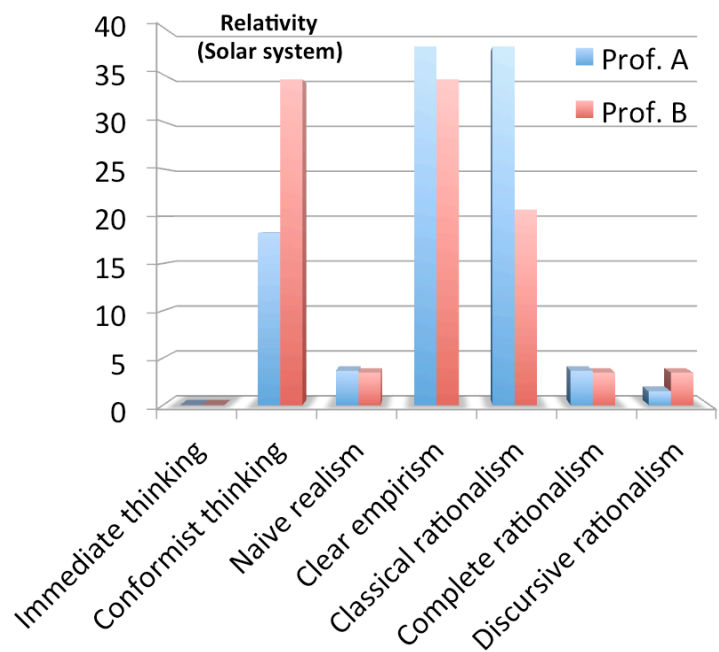

FIG. 8. Same as Fig.7, but now from an astronomical point of view

## 4. Some modalities of resurgence of conformist thinking

After attending a class for primary school teachers ending with a discussion of the equal legitimacy of an observer's point of view on the Earth and that of an observer on the Sun [15, 49], Prof. A strongly disagreed on the 'equivalence' between terrestrial and solar points of view. An epistolary exchange was then started by sending him excerpts from Mach's *Science of Mechanics* (1883 [42]). A constant during these exchanges was the search by Professor A of experiences or facts that might justify the non-equivalence of the two reference systems: '*Is there not a way to show that the two points of view are not equivalent?*' Even though he easily admits in oral discussions that the very impossibility of such means is indeed one of the basic principles of the general theory of relativity (1916), he nevertheless argues that some models would be '*truer*' than others, because of a supposedly better 'explanatory power', even if he recognizes their formal equivalence and offers no theoretical argument in favor of such a distinction. Einsteinian when it comes to the principles or problems explicit involving clocks, Prof. A becomes fiercely Newtonian again when it comes to reasoning concretely on a geocentric model. This modality seems largely unconscious in his case.

Prof. B expressed even more strongly her disagreement at the same course, stating authoritatively that '*there are plenty of experiences that show that the Earth is spinning [around the Sun]*'. When asked to quote at least one, she changes her argument and shifts into a highly passionate mode: '*You can not say things like that to teachers, because otherwise everything is relative and you open the way to all the excesses. We can't jeopardize all the work we've done to make it clear that the Earth is turning.*' She fully assumes to encourage teachers to wrongfully condemn an actually legitimate interpretation, '*Let [the students] say that it is false rather than letting them believe that the Sun turns around Earth, because they would not understand,*' and considers that in this situation '*it is not necessarily welcome to shake the teachers' views.*'

She thus expresses, in a fully conscious and deliberate way, though very passionate, a posture of thought resolutely conform to an obsolete, though dominant, Copernican paradigm and in contradiction not only with her own scientific knowledge, but also with the primacy of experience, the latter being in his opinion secondary in this specific case. The high importance that she attributes *as a physics teacher* to the heliocentric approach, and therefore the strong conformist component of her response, seems to be much less a question of her own adherence to this model (she claims to be fully Einsteinian *as a researcher*) than of the mission that she considers as essential and a priority, as a representative of the scholarly community, to block ideas that she considers dangerous for unprepared minds, such as primary school teachers, let alone their students. One may be tempted to recognize there, at this time, a sociological posture of 'gate-keeper' [41] regulating and limiting the expression of these ideas outside the authorized circle of the recognized physics community, possibly reinforced in the case of relativity by the idea that 'the public is irretrievably doomed to incomprehension' being, according to French sociologist Pierre Bourdieu, '*so profoundly embedded in the social definition of the intellectual's vocation that it tends to be taken for granted*' [48, 50].

*5. Synthesis*

For physicist Prof. A, a moderate paradigmatic pressure results in a noticeable increase in conformist thinking (Fig. 6, velocity of light). A passage to a strong paradigmatic pressure (geocentrism, (Fig. 8, left) further increases this component, which nevertheless remains rather minor, since a kind of mechanism of epistemological self-defense seems to set up with a relative increase in the empiricist component at the expense of rationalisms complete and discursive (Fig. 9).

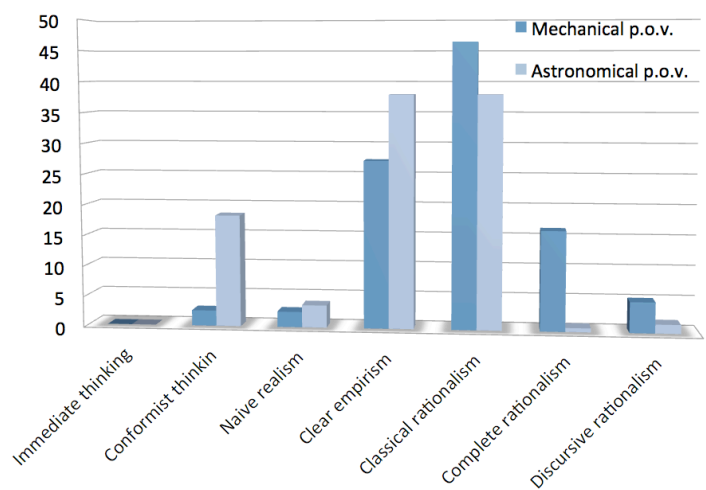

FIG. 9. Significant change in Prof. A's epistemological profile relative to the concept of relativity of motion between the low (mechanical) and strong (astronomical) pressure point of views

A moderate paradigmatic pressure does not induce a significant increase in conformist thinking in Prof. B, also a physicist, while the discursive rationalist component increases a little (Fig. 5, right). On the other hand, a disturbing approach to the Copernican problematic makes her switch, no doubt deliberately, towards a largely unscientific gate-keeper posture, including a strong conformist component (Fig. 10).

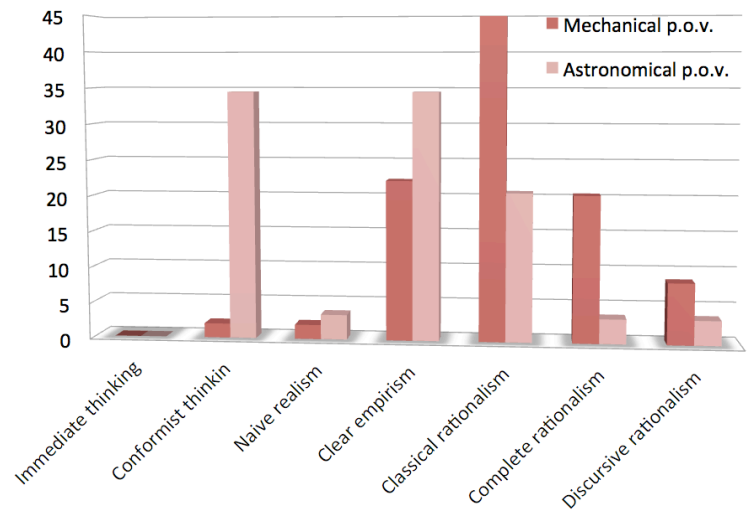

FIG. 10. Same as Fig. 9, but for Prof. B

In both cases, each physicist's epistemological profiles relative to the concepts of velocity of light (Figs. 4 & 5) as well as to the concept of relativity of motion (mechanical point of view, Fig. 7) remain fairly similar (despite a stronger 'classical rationalism' component in the latter) as long as only no formulation likely to induce a strong paradigmatic pressure is involved. On the other hand, we observe a dramatic change of both profiles relative to the concept of relativity of motion as soon as the formulation involves the Copernican issue (Figs. 9 & 10, right). It should be noted that the empiricist and classical rationalist components always remain significant, even when some conformist thinking sets up.

### B. A purely mathematical response

While informal discussions on other physical concepts suggest for the other, non-physicist academics of the sample profiles also dominated by the empiricist and rationalist components, more detailed exchanges on the heliocentrism problem reveal more singular profiles to this respect.

Figure 11 presents the expanded epistemological profiles of Prof. C, a professional mathematician, relative to the concept of relativity of motion considered from an astronomical point of view.

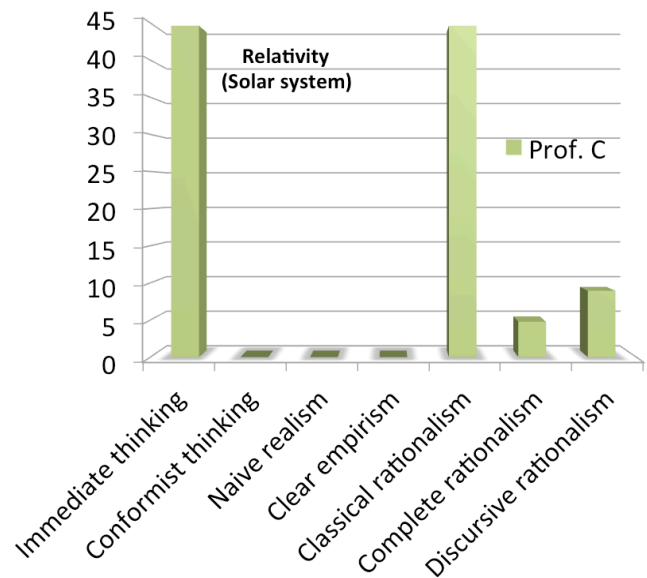

FIG. 11. Expanded epistemological profile of Prof. C relative to the concept of relativity of motion formulated form an astronomical point of view (arbitrary units). Note the strong 'Immediate thinking' component

Prof. C claims a good mastery of Newtonian celestial mechanics but, rather than historical advances in physics (Galilean or Einsteinian), the equal legitimacy of the two reference frames appears to be a mathematical banality for him, based on arguments of patent symmetry. This a-physical posture is an unexpected form of thought that is both immediate and scholarly, with no identifiable conformist component, even in the form of a social norm [6].

It is interesting to note that Prof. C is also a regular reader of 'hard science ction', including novels speculating about alternate theories of Relativity, such as Greg Egan's *Orthogonal* trilogy [51], and likes to speculate about the corresponding formalisms, which may account for the significant 'discursive rationalism' of his profile relative to this notion.

Generally, he seems fairly insensitive to the paradigmatic pressure associated with physical questions.

## C. Effect on a Sociologist

### *1. Short-term response*

Also claiming a certain (although possibly more questionable) mastery of Newtonian mechanics, Dr. D is at first surprised by the relativistic argument, that he says to be discovering. He nevertheless seems to understand and accept it, though perhaps partly on the basis of the authority he recognizes in the matter to his interlocutors.

His epistemological profile presents a strong conformist component, together with a significant rationalist one but no, or only very weak, empiricist component (Figure 12).

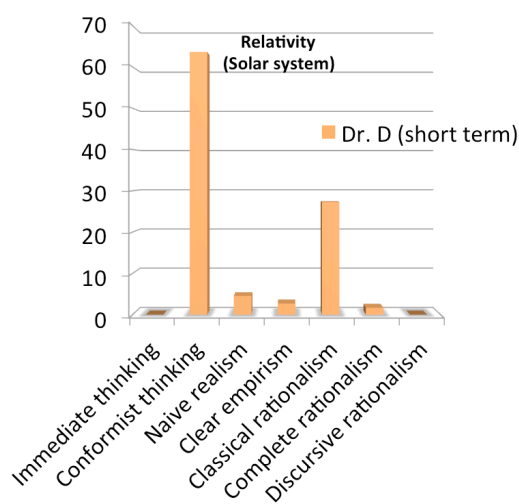

FIG. 12. Short-term expanded epistemological profile of Dr. D relative to the concept of relativity of motion formulated from an astronomical point of view (arbitrary units)

### *2. Delayed response*

A few days after the interview, after he had conversations with other non-specialist teachers, he clearly receded this acquiescence, by mail, on the basis of arguments which he himself recognizes as 'ideological'. He explained that '*geocentrism is associated with clerical obscurantism and the heliocentric system marks the victory of scientific audacity against religious dogma. Of course, these 'political' biases are part of the epistemological obstacle*' [also a Bachelardian concept [8], discussed with Dr. D during his interview]. He clearly identifies a problematic relationship with authority and insists that the Einstein quote: '*The two propositions, 'the Sun is motionless and the Earth rotates' or 'the Sun turns and the Earth is motionless', merely signify two different conventions concerning two different coordinate systems.*' [52], although it '*seems logical to [him]*', actually induced in him a semi-conscious attitude of the kind: *'Keep talking, but I'm not listening'...*' ['*Cause toujours, je n'écoute pas*' — his phrasing]

### *3. Long-term response*

On his next meeting with one of the authors (EB), more than a year after these exchanges, Dr. D came to her on his own initiative to tell her that, after some time had passed, he had decided to read about the theories of relativity, including some original works of Galileo [13] and Einstein [52] themselves but also from other authors, had attended popularization conferences, etc. He could not remember a particular episode that had him change his mind again, but the principle of relativity did now appear to him '*just obvious*'. Another meeting, a few months later, confirmed that his position seemed serenely stabilized.

It was not possible to set up a new interview with the two authors, but it seems reasonable to infer that Dr. D had fully transitioned from a at least partial adherence to the Copernican heliocentric paradigm to a full participation to the Galilean, or even Galileo-Einsteinian, relativist paradigm.

Also an avid reader of hard science fiction, Dr. D did not seem to seriously question the concept of relativity in his new understanding (discursive rationalism), nor did attempt to relativize them with regard to even more sophisticated approaches, such as General relativity (complete rationalism), and still he makes no direct reference to experimental evidences (empiricism).

If, regardless of the difference in procedure, we try to translate these observations onto a new epistemological profile of Dr. D and compare it to the earlier one (Fig. 13), it appears that the profile remained mostly unchanged: while its 'conformist thinking' component has dramatically shifted from compliance to the Copernican paradigm towards compliance to the Galilean paradigm, it still remains a key feature of his profile.

Our sociologist appears extremely sensitive to Copernican paradigmatic pressure and shifts in full awareness towards a largely conformist thought. It is interesting to note that it is still in full awareness to comply to the dominant paradigm that he shifts again towards the relativist paradigm, after making sure from several different sources that it is indeed the current consensus among physicists.

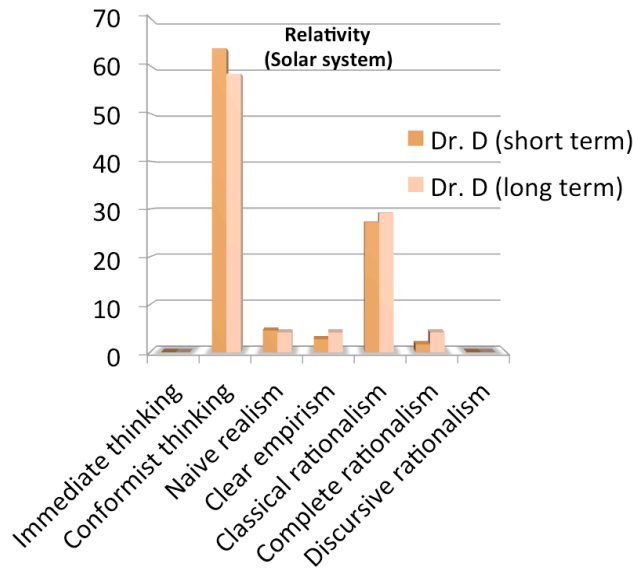

FIG. 13. Expanded epistemological profiles of Dr. D relative to the concept of relativity of motion formulated form an astronomical point of view before (left, darker) and after (right, lighter) his final shift towards the relativist paradigm. While conformist thinking does is no longer complies to the same paradigm, the aspect of the epistemological profile remains mostly unchanged

### D. Comparison and discussion

*1. Specialty-dependence of epistemological profiles*

Despite the very limited size of our sample, an obvious feature of the considered epistemological profiles is the general similarity of all physicist's profiles (Profs A & B, Figs. 4-10), at least as far as the conformist component is not considered, while they have very little in common either with a mathematician's (Prof. C, Fig. 11) or a sociologist's (Dr. D, Figs. 12 & 13) profiles. It must nevertheless be kept in mind that all the considered profiles were relative to physical notions, and that these differences likely derive from comparing specialists and non-specialists profiles.

The strong and robust conformist component of Dr. D's profiles relative to physical notions thus doesn't suggest in any way that conformist thinking is a normal component of sociology. Much to the contrary, one can speculate that this component might well totally disappear in his profiles relative to the sociological notions he is a specialist of, while a symmetric conformist component might appear in the physicists' profiles relative to these sociological notions. Unfortunately, the varication of this intuition is well beyond the scope of this study, as the authors definitely lack the sociological hindsight that would be necessary to establish such profiles.

Similarly, it would be interesting to determine whether the apparent insensitivity of Prof. C to the pressure associated to physical paradigms remains in his own field, and to study his response to the strong pressure associated with a disturbing approach to a mathematical paradigm.

*2. A nonlinear response?*

Dr. D's brutal changes in epistemological profiles, both short- and long-term, followed by periods of stability, is somewhat reminiscent of the behavior of some bistable systems, already observed in opinion dynamics [53]. Since Prof. B rapid shift from strictly scientific to 'gate-keeper' postures might also be reminiscent of bifurcation-type behaviors, these observations suggest that individual responses to paradigmatic pressure might be highly non-linear processes depending on the pressure level, and therefore probably largely unpredictable for a given individual, even if it was possible to quantify precisely this level. On the other hand, such nonlinear models could have a statistical sense about large populations, in the spirit of sociophysics [54].

*3. Features of scientific conformity*

Our observations strongly suggest several striking features of conformist thinking relative to scientific issues:

- Scar components of conformist thinking remain latent even in highly informed and professional scientific minds.

Indeed, although any generalization from our limited data would be abusive, they are in full accordance with Gaston Bachelard's own assumptions and conclusions about the formation of the scientific mind [8], and his very motivation for developing epistemological profiles as a tool to take such components into account [34], even if he didn't explicitly consider the case of conformist thinking.

- Strong paradigmatic pressure appears to be an efficient trigger for conformist thinking.

- The emergence of conformist thinking can either happen together with the simultaneous emergence of the complete & discursive rationalist components or, to the contrary, at their expense
- Under strong paradigmatic pressure, the emergence of conformist thinking may follow very different rationale.

The last two items might be strongly correlated, as exemplified by the very different responses of Profs. A & B (Figs. 9 & 10). Finally:

- An epistemological profile involving a strong conformist component may remain mostly unchanged even when the bearer's understanding of the considered notion undergoes a complete reversal.

In other words, a conformist thinking process relative to a given notion may appear mostly independent from the actual paradigm the bearer is conforming to (namely, Copenican or Galilean as exemplified by the case of Dr. D, Fig. 13).

# VI. ALTERNATIVE TEACHING STRATEGIES TO AVOID CONFORMIST THINKING

## A. Daring historical approaches

In terms of science education, a key issue deriving from the above findings is that a same given scientific concept (e.g. the principle of the relativity of motion) can appear either fairly easy or extremely difficult to teach efficiently, and for the students to accept, depending on the level of paradigmatic pressure associated, not necessarily to the concept itself, but to some formulations or examples likely to trigger it.

A direct corollary seems to be that, as far as the transmission of scientific concepts only is concerned (as opposed to the teaching of the history of science), teachers might find it advantageous to avoid 'hot' issues and formulations, more likely associated with stronger paradigmatic pressures, and on the contrary to favor harmless examples. For instance, teaching the physics of Galilean relativity though considerations over merry-go-rounds rather than the relative motion of the Earth and the Sun...

Yet, as far as the associated paradigmatic pressure can be an indication of the importance of a given issue in our common scientific culture, some teachers may also consider it part of their duty to have their students confront some hot issues.

If we admit that the efficiency of direct, head-on strategies is likely to be very low, as suggested by the disturbingly high proportion of conformist Copernican thinking observed even among rather advanced students, are other strategies available ?

While some scientists can appear rather pessimistic, like Max Planck who is quoted by Thomas Kuhn [38] saying that '*a new scientific truth does not triumph by convincing its opponents and making them see the light, but rather because its opponents eventually die, and a new generation grows up that is familiar with it*', others, including the very pioneers of modern science, attempted to develop new transmission strategies when confronted with the (then presumably even stronger than today!) paradigmatic pressure (then Ptolemean rather than Copernican, obviously) associated with the heliocentric concept.

While Giordano Bruno's and Galileo's efforts brought such literary and historical gems as *The Ash Wednesday Supper* [55] or *The Starry Messenger* [56], the most interesting enterprise might be Johannes Kepler's. As a student, and as early as 1593, Kepler devoted one of his required dissertations to the question: '*How would the phenomena occurring in the heavens appear to an observer stationed on the Moon?*' [57]. Later, as a full-edged astronomer, his key work in defence of the Copernican system, the *Astronomia Nova* [58], established that, whatever the point of view, observation alone does not allow to determine whether a celestial body is fixed or not — hence to decide between the Copernican and the Ptolemean systems. However, regardless of its importance fo the history of science, the *Astronomia Nova* is a highly mathematized work, inabordable to the common reader, and possibly even to most early-XVII$^{th}$ Century astronomers.

Kepler then went back to the narrative approach, and to the idea of a Lunar observer, with the *Somnium* [59] —which some argue might also be considered the very first science fiction novel [60]. Despite being a major literary, and personal, failure [57] (written ca. 1609, it only appeared as a posthumous work in 1634), the *Somnium* was a brilliant and explicit attempt to reach and convert a wider audience. Getting the reader to suspend his incredulity and accept, for the duration of a story, to partake to a Lunar point of view allowed him avoid the (then) considerable pressure of the Ptolemean paradigm, focused on the Terran and Solar points of view. Then, the reader's acceptance of the legitimacy a particular non-Terran (namely, Lunar) point of view would hopefully allow him to also accept the legitimacy of other non-Terran points of view, such as a Solar one — and thus easily shift to heliocentrism.

Similarly, it has been argued that accepting the concept of earthshine, introduced by Galileo in his watercolors of the Moon [56, 61] would also have been seen '*as supporting a Copernican world system*' [62].

Taking advantage of modern fast computing techniques to instantly shift the point of view of spectators immersed in a planetarium from geocentric to allocentric, for instance selenocentric, has also been proposed to help participants understand the phases of the Moon [63].

Other original approaches include theatrical reenacting of Galileo's *Dialogue* [13, 64].

### B. Two IBSE sequences

In Kepler's spirit, an investigation-based science education sequence, '*Will Earth come out of the frame?*' [49] was designed around a painting by science fiction artist Manchu, '*Framed Earth*' (Fig.14) [65] to have the participant work on the idea of an equal legitimacy of all reference frames.

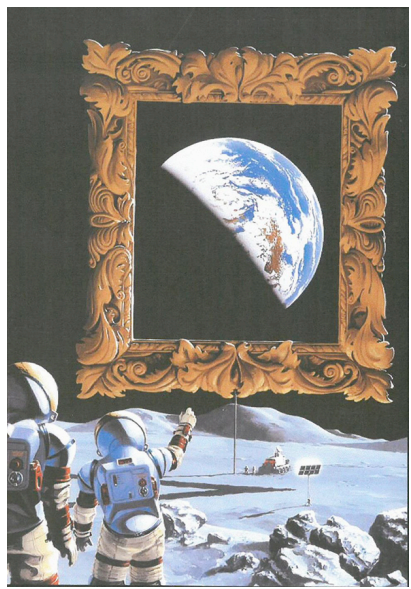

FIG. 14. 'Framed Earth' by Science Fiction artist Manchu

Another sequence had them work on the spatial geometry of orbiting frames by having them represent the Earth and the Earth with their bodies, with one student 'orbiting' around the other, or manipulating Styrofoam spheres (Fig.15). The latter sequence is rather classical, and adapted to primary school pupils and their teachers, and generally to participants of any age not fully at ease with the relevant mathematical tools or not having good enough spontaneous spatial visualization [66] (like presumably most future teachers).

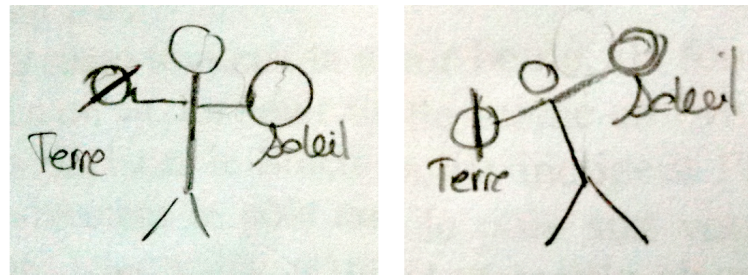

FIG. 15. Future primary school teachers drawings of themselves holding spheres representing the relative motion of the Earth (Terre, in French) and the Sun (Soleil) (seq. 'Spatial Geometry')

### C. Necessity of a dual strategy

A group of 93 post-graduate students in a primary school teaching track was thus subjected to the questionary described in Section II, in its modified Formulation 4 version: *'Two pupils are talking. One asserts that the Earth revolves around the Sun, the other that the Sun revolves around the Earth. They turn to you for your opinion. What do you answer them?'* They worked first on the 'Framed Earth' sequence, then took the questionnaire. With almost 90% of the answers of students not having previously followed the full course qualifying as Copernican, this approach appears almost entirely inefficient by itself (while, by contrast, 100% of the redoubling students qualified as Galilean).

They then participated to the 'spatial geometry' sequence, and took the questionnaire again. This time, 60% qualified as Galilean (Table VIII) . The combination of the two sequences combining a training of the ability to change points of view with a work on the equal legitimacy of all points of views thus appear far more efficient.

|  | Number | Copernican | Galilean |
|---|---|---|---|
| After sequence 'Framed Earth' only | 85 | 82 (96.5 %) | 3 (3.5%) |
| After both sequences | 15 | 18 (40%) | 27 (60%) |

TABLE VIII. Proportion of students validating ('Galilean' answer) or not ('Copernican' answer) Formulation 4 depending on the sequence(s) they participated to.

Subsequent discussion showed that many of these students had initially failed to even identify the problem, however ancient and conceptually simple it may appear to some others and in spite of having repeatedly studied it both in primary and secondary school. By itself, the usual basic strategy of directly training the participants to shift points of view (or reference frames, for the more mathematically inclined, as Kepler attempted in his *Astronomia Nova* [58]) thus appears fairly inefficient.

## VII. SYNTHESIS AND CONCLUSION

In the first part of this article, we have established that, while very real in certain contexts, some common faulty answers to basic astronomy questions, often interpreted as a sign of poor scientific literacy, can often be triggered by the formulation of the question. Indeed, a formulation focused on some aspects of a given paradigm, either current or obsolete, may induce a strong paradigmatic pressure and dramatically modify a person's epistemological profile relative to the underlying physical notions.

Yielding to such a pressure appears to be a very common cognitive process among the general public, to the point that, whether or not they're familiar with the concepts of Galilean relativity, an overwhelming proportion of students, as well as of teachers and future teachers, fall back into an obsolete, pre-Galilean thinking, namely conformation to Copernican heliocentrism.

Indeed, the paradigmatic pressure associated with the Copernican paradigm remains very high for certain formulations, and may profoundly alter people's epistemological profiles.

For the same physical concept, the paradigmatic pressure can be very different depending on the triggering situation. Thus, the pressure associated with the concept of relativity, which appears to be very strong for triggering situations in resonance with the very old Copernican quarrel, is on the contrary quite moderate for more neutral issues (e.g. relative motions of trains or merry-go-rounds).

The effects of paradigmatic pressure on epistemological profiles are quite diverse. The most obvious is the appearance or increase of conformist thinking with the increase of the paradigmatic pressure.

A direct corollary in science education is the necessity for teachers to be aware of the paradigmatic pressure associated to certain issues or formulation. Then, they can either deliberately avoid them and systematically prefer lower-pressure formulations, less prone to trigger conformist thinking instead of more scientific empiricist or rationalist postures (or combinations thereof) among the students; or, when they find that they can not avoid confronting a triggering issue, use carefully designed sequences, or sequence combinations, to release some of the pressure before bringing up the problematic issues or formulations. They can for instance, in the spirit of Kepler's *Somnium*, make a detour to deal with the legitimacy of a Lunar point of view before considering that of a (strongly triggering in astronomy) geocentric point of view in a relativistic approach.

We have also shown that even very sophisticated and informed minds, such as those of veteran physicists, fully proficient in the theories of relativity, are not immune to such a pressure, and can also fall back into conformist thinking to pre-relativist paradigms.

We wish to emphasize that, despite the authors' avowed preference for making the more modern ideas of relativity available to teachers and students whenever possible, this study does not, in itself, invalidate the more cautious, gate-keeper-like arguments expresse for instance by Prof. B [cf. *§V.A.4.*], the importance of which we also fully acknowledge. A more thorough study on the long term effects of the teaching of Galilean relativity in primary school will be necessary to decide this debate.

Finally, the epistemological profiles developed by Gaston Bachelard has proven to be well adapted to the evaluation of the paradigmatic pressure associated with a particular approach of a given concept, for a cultivated public and able to master the technical subtleties. Further studies will also be necessary to determine if a (at least statistically) predictive nonlinear model can be developed to describe the response of a given profile to a strong paradigmatic pressure.

## Acknowledgements

The authors wish to thank all participants to this study, and especially Profs A, B, & C and Dr. D for their kindness, availability and sincerity. EB also thanks Yves Gingras for a useful discussion and Daniel Hennequin for his help in the organization of the SFP questionnaire.

**References**

[1] Youtube, excerpt of the TV game *Qui veut gagner des millions?* (France, 2015). Online: https://www.youtube.com/watch?v=ekmtqODjrSI (retrieved Nov. 22, 2019)

[2] Steve Crabtree, New Poll Jauges Americans' Knowledge Levels (1999). From Gallup website: http://www.gallup.com (retrieved Nov. 22, 2019)

[3] National Science Fundation, Science and Engineering Indicators 2014, ch. 7 (2014). From NSF website: https://www.nsf.gov/statistics/seind14/index.cfm/chapter-7/c7h.htm (retrieved Nov. 22, 2019).

[4] Andreas Flache, Michael Mäas, Thomas Feliciani, Edmund Chattoe-Brown, Guillaume Deuant, Sylvie Huet, and Jan Lorenz, Models of Social Influence: Towards the Next Frontiers, Journal of Artificial Societies and Social Simulation, **20**, 4, p. 2 (2017).

[5] Solomon E. Asch, Effects of group pressure upon the modification and distortion of judgments. In H. Guetzkow (ed.), *Groups, leadership and men; research in human relations,* Oxford, England: Carnegie Press, p. 177-190 (1951).

[6] F. Maijn Stok and Denise T.D. de Rider, The Focus Theory of Normative Conduct, in *Social Psychology in Action*, K. Sassenberg, M.L.W. Vliek (eds.), Springer Nature Switzerland, p. 95-110 (2019).

[7] Robert B. Cialdini and Noah J. Goldstein, Social Influence: Compliance and Conformity, Annu. Rev. Psychol., **55**, p. 591-621 (2004).

[8] Gaston Bachelard, *The Formation of the Scientific Mind* (*La Formation de l'esprit scientifique*, 1938); Clinamen Press, 2006.

[9] William C. Wimsatt, *Re-engineering Philosophy for Limited Beings: Piecewise Approximations to Reality,* Cambridge, Mass.:Harvard Univ. Press (2007).

[10] Donald T. Campbell, Blind variation and selective retentions in creative thought as in other knowledge processes, Psychological Review, **67**, 6, p. 380-400 (1960).

[11] Thomas Kuhn, *The Copernican Revolution: Planetary Astronomy in the development of Western Thought* (1957); Cambridge, Mass.:Harvard Univ. Press (2007).

[12] Nicolaeus Copernicus, *On The Revolutions of the Celestial Spheres* (*De revolutionibus orbium coelestium*, 1543); Prometheus Books, 1995.

[13] Galileo Galilei, *Dialogue Concerning the Two Chief World Systems* (*Dialogo sopra i due massimi sistemi del mondo*, 1632), Modern Library, 2001.

[14] Henri Poincaré, *Science and Hypothesis* (*La Science et l'hypothèse*, 1902); Cornell Un. Library, 2009.

[15] Albert Einstein, À propos de *La Déduction relativiste* d'É. Meyerson, La Revue philosophique, **CV**, p. 161-166 (1928).

[16] Kevin J. Pugh, Mark Girod, Science, Art, and Experience: Constructing a Science Pedagogy From Dewey's Aesthetics, J. Science Teacher Educ., **18**, p. 9-27 (2007) .

[17] Robert S. Westman, *The Copernican Question. Prognostication, Skepticism, and Celestial Order*, Berkeley:Univ. California Press (2011).

[18] Dennis Danielson, The Great Copernican Cliche, Am. Journal Physics, **69**, 10, p.1029-1035 (2001).

[19] Yves Gingras, personnal communication.

[20] Ricardo Trumper, Teaching Future Teachers Basic Astronomy Concepts— Seasonal Changeslat a Time of Reform in Science Education, J. Res. Science Teaching, **43**, 9, p. 879-906 (2006).

[21] Cécile de Hosson, Isabelle Kermen, and Étienne Parizot, Exploring students' understanding of reference frames and time in Galilean and special relativity, Europ. J. Phys., **31**, 6, p. 1527 (2010).

[22] Jennifer Wilhelm, Merryn Cole, Cheryl Cohen, and Rebecca Lindell, How middle level science teachers visualize and translate motion, scale, and geometric space of the Earth-Moon-Sun system with their students, Phys. Rev. Phys. Educ. Res., **14**, 010150 (2018).

[23] Alessandro Albanese, Marcos Cesar Danhoni Neves, and Matilde Vicentini, Models in science and in education: A critical review on research on students' ideas about the Earth and its place in the universe, Science & Education, **6**, p. 573-590 (1997).

[24] Édith Saltiel, Spontaneous ways of reasoning in elementary kinematics, Europ. J. Physics, **1**, p.73-80 (1980).

[25] Sudhir Panse, Jayashree Ramadas, and Arvind Kumar, Alternative conceptions in Galilean relativity: Frames of Reference, Int. J. Science Educ., **16**, 1, p. 63-82 (1994).

[26] Jayashree Ramadas, Shrish Barve, and Arvind Kumar, Alternative conceptions in Galilean relativity: inertial and non-inertial observers, Int. J. Science Educ., **18**, 5, p. 615-629 (1996).

[27] J.S. Aslanides, and Craig M. Savage, Relativity concept inventory: Development, analysis and results, Phys. Rev. ST - Phys. Educ. Res., **9**, 010118 (2013).

[28] Ramadas, Jayashree, Barve Shrish, and Kumar, Arvind, Alternative conceptions in Galilean relativity: distance, time, energy and laws, International Journal of Science Education, **18**, 4, p. 463-478 (1995).

[29] Atanu Bandyopadhyai, Students's ideas of the meaning of the relativity principle Europ. J. Physics, **30**, p.1239-1256 (2009).

[30] Estelle Blanquet and Daniel Hennequin, La Culture scientifique des bacheliers, General Congress of the Société Française de Physique, Orsay, France, July 3-7 (2017).

[31] B.O. (2010), Programme de Physique-Chimie en classe de seconde générale et technologique, Bulletin Officiel spécial 4 du 29 avril 2010.

[32] Christian S. Crandall, Amy Eshleman, and Laurie O'Brien, Social norms and the expression and suppression of prejudice: the struggle for internalization, J. Personal. Soc. Psychol., **82**, p. 359-378 (2002).

[33] Immanuel Kant, *Critique of Pure Reason* (*Kritik der reinen Vernunft*, 1781), Penguin Classics, 2008.

[34] Gaston Bachelard, *The Philosophy of No: A Philosophy of the New Scientific Mind* (*La Philosophie du non*, 1940), Orion Press, New York, 1968.

[35] Estelle Blanquet and Éric Picholle, Profil épistémologique de 777 enseignants français du primaire au regard de la science, in *Actualités et perspectives des recherches en didactique des sciences et des technologies*, I. Kermen ed., France:Presses Un. Artois, p. 44-56 (2018).

[36] Karl Popper, *The Logic of Scientific Discovery*, Routledge Classics (1934).

[37] Paul Feyerabend, *Against Method* (1975), Verso Pub, 2010.

[38] Thomas Samuel Kuhn, *The Structure of Scientific Revolutions* (1962), Univ. Chicago Press, 2012.

[39] E.Margaret Evans, Cristine H. Legare, and Karl S. Rosengren, Engaging multiple epistemologies, in *Epistemology and Science Education: Understanding the Evolution vs. Intelligent Design Controversy*, R. Taylor & M. Ferrari eds., London:Routledge, p. 111-139 (2011).

[40] Kurt Lewin, Forces behind food habits and methods of change, Bulletin of the National Research Council, **108**, p. 35-65 (1943).

[41] Karine Barzilai-Nahon, Toward a Theory of Network Gatekeeping: A Framework for Exploring Information Control, Journal of the American Information Science and Technology, **59**, 9, p.1-20 (2008).

[42] Ernst Mach, *The Science of Mechanics* (*Die Mechanik in ihrer Entwickelung : historisch-kritisch dargestellt*, 1883), Open court, 1942.

[43] Léon Brillouin, *Wave Propagation and Group Velocity*, Academic Press, New York, 1960.

[44] Éric Picholle and Carlos Montes, Un exemple récent de réduction de pression paradigmatique: l'évolution de la réception de la notion de vitesse supraluminique, 1$^e$ Rencontres Optique et Didactique (REOD), Limoges, France (2017).

[45] Éric Picholle, Carlos Montes, Claude Leycuras, Olivier Legrand, and Jean Botineau, Observation of dissipative superluminous solitons in a Brillouin fiber ring laser, Phys. Rev. Lett., **66**, p. 1454-1457 (1991).

[46] Zhongyang Li, Zizhuo Liu, and Koray Aydin, Wideband zero-index metacrystal with high transmission at visible frequencies, J. Opt. Soc. Am B, **34**, 7, p. D13-D17 (2017).

[47] Jamie Arndt, Je Schimel, Je Greenberg, and Tom Pyszczynski, The Intrinsic self and defensiveness: evidence that activating the intrinsic self reduces selfhandicapping and conformity, Personal. Soc. Psychol. Bull., **28**, p. 671-683 (2002).

[48] Thomas F. Glick, Cultural Issues and Relativity, in *The Comparative Reception of Relativity,* Dordrecht:D. Reidel Pub., 1987, p. 381-400.

[49] Estelle Blanquet, Astronomie et mouvement relatif: sortir du cadre, in *Science et fictions à l'école: un outil transdisciplinaire pour l'investigation?*, Nice, France:Somnium, p. 149-177 (2011) [in French].

[50] Pierre Bourdieu, Intellectual Fields and Creative Projects, in *Knowledge and Control*, M.D. Young ed., London:Collier-Macmillan, 1971, p. 161-188.

[51] Greg Egan, *The Orthogonal Trilogy : The Clockwork Rocket* (2011); *The Eternal Flame* (2012); *The Arrows of Time* (2013), San Francisco:Night Shade Books.

[52] Albert Einstein and Leopold Infeld, *The Evolution of Physics: From Early Concepts to Relativity and Quanta* (1938), Touchstone, (2008).

[53] Shaoli Wang, Libin Rong, and Jianhong Wu, Bistability and multistability in opinion dynamics models, Applied Mathematics and Computation, **289**, p. 388-395 (2016).

[54] Serge Galam, Modeling Radicalization Phenomena in Heterogeneous Populations, PLos ONE, **11**, 5, p. 1-15 (May 2016).

[55] Giordano Bruno, *The Ash Wednesday Supper* (*La Cena de le Ceneri*, 1584), Toronto, Canada:Univ. Toronto Press (2018).

[56] Galileo Galilei, *The Starry Messenger* (*Sidereus Nuncius*, 1610), Levenger Pub. (2013).

[57] James R. Voelkel, *The Development and Reception of Kepler's Physical Astronomy, 1593-1609*, Indiana Un. Press 1994).

[58] Johannes Kepler, *New Astronomy* (*Astronomia Nova*, 1609), Green Lion Press, 2015.

[59] Johannes Kepler, *Somnium, or Posthumous Works on Lunar Astronomy* (*Somnium, seu opus posthumum de astronomia*, 1634), Dover Publications, 2003.

[60] Gale E. Christianson, Kepler's Somnium: Science Fiction and the Renaissance Scientist, Science Fiction Studies, **3**, 1 (1976).

[61] Eileen Reeves, *Painting the Heavens: Art and Science in the time of Galileo*. Princeton:Princeton Univ. Press (1997).

[62] Paolo Molaro, On The Earthshine Depicted in Galileo's Watercolors of the Moon. Galilaeana, X, p. 73-84 (2013).

[63] Pierre Chastenay, From Geocentrism to Allocentrism: Teaching the Phases of the Moon in a Digital Full-Dome Planetarium, Research in Science Education, **46**, 1, p. 43-77 (2015).

[64] Vasilis Tselfes and Antigoni Paroussi, Science and Theatre Education: A cross-disciplinary approach of scientific ideas adressed to student teachers of early childhood education, Sci. & Educ., **18**, p. 1115-1134 (2009).

[65] Manchu, *Science [Fiction],* artbook, Paris:Delcourt pub. (2004).

[66] Maria Kozhevnokov, Michael A. Motes, and Mary Hegarty, Spatial visualization in physics problem solving, Cognitive Science, **31**, 4, p. 549-579 (2007).